\documentclass[aps, prb, twocolumn, superscriptaddress, longbibliography]{revtex4-1}
\usepackage{amsmath,amssymb}
\usepackage{graphicx}
\usepackage{dcolumn}
\usepackage{bm}
\usepackage{color} 
\usepackage{multirow}
\usepackage{siunitx}
\usepackage[colorlinks = true, urlcolor = blue, linkcolor = blue, citecolor=blue]{hyperref}

\begin{document} 
\title{Magnetic transitions of biphenylene network layers induced by external perturbations}
\author{Sejoong Kim}
 \email{sejoong@alum.mit.edu}
 \affiliation{University of Science and Technology (UST), Gajeong-ro 217, Daejeon 34113, Korea}
 \affiliation{Korea Institute for Advanced Study, Hoegiro 85, Seoul 02455, Korea} 

\date{\today}

\begin{abstract}
We present a comprehensive investigation of the magnetic ordering in biphenylene network (BPN) layers, employing density functional theory (DFT) calculations under external perturbations, including uniaxial strains and hole doping. We compute fully relaxed structures, energy bands, and magnetic states by performing DFT calculations augmented with extended Hubbard interactions, encompassing both on-site and inter-site interactions, to accurately capture electron correlations. 
We emphasize the importance of the extended Hubbard forces by contrasting BPN layers with and without the forces. 
Our results reveal that in their fully relaxed structures, both BPN monolayer and bilayer are non-magnetic. 
We exploit external perturbations to induce magnetic ordering. 
The application of uniaxial strains induces magnetic phase transitions, leading to ferrimagnetic and antiferromagnetic states in BPN monolayer and bilayer, respectively. Additionally, we investigate hole doping as an alternative mechanism for inducing magnetic transitions. 
Our findings shed light on the tunability of magnetic properties in BPN layers through external perturbations, demonstrating the promise of low-dimensional materials in future spintronics and nanoelectronic applications.
\end{abstract}

\maketitle

\section{\label{sec:Intro}Introduction}
Recently a new planar carbon allotrope has been experimentally synthesized on the Au(111) surface throughout a bottom-up process~\cite{doi:10.1126/science.abg4509.Fan}. 
The carbon allotrope is called biphenlyene network (BPN), which consists of octagon, hexagon, and tetragon rings of $sp^2$-hybridized carbon atoms. 
In the experiment BPN was grown as an armchair-edged nanoribbon structure whose width is up to 18 carbon atoms, but it is certainly expected to realize a large-scale two-dimensional BPN single-layer in the future. It would serve as a versatile platform to investigate fundamental physics of condensed matter systems as graphene has done so far~\cite{RevModPhys.81.109}. 

The experimental realization of BPN layers has sparked significant interests in exploring their electronic, phononic, mechanical, chemical, and thermal properties~\cite{doi:10.1021/acs.nanolett.2c00528.Son, PhysRevB.104.235422, 10.1063/5.0102085, HAMEDMASHHADZADEH2022111761, MORTAZAVI2022100347, D1NR07959J, REN2023112119, PhysRevB.105.035408, Luo2021, Bafekry_2022, https://doi.org/10.1002/jcc.26854, LIU2022153993, CHOWDHURY2022110909, doi:10.1021/acsaelm.2c00459, doi:10.1021/jacs.2c02178, D1CP04481H, D2RA03673H, 10.1063/5.0088033, Zhang_2022, Ren_2022, Ge_2022, D2CP04381E, VEERAVENKATA2021893, D1TC04154A, https://doi.org/10.1002/aenm.202200657, LI2022349, XIE2023112041, 10.1063/5.0140014, doi:10.1021/acs.jpclett.1c03851, https://doi.org/10.1002/est2.377, Asadi2022, SU2022108897, Al-Jayyousi2022, D2CP00798C, D2CP04752G, D2NH00528J}. 
Recent studies~\cite{doi:10.1021/acs.nanolett.2c00528.Son, PhysRevB.104.235422} have revealed that the 2D BPN lattice exhibits intriguing electronic band structures such as type-II Dirac cones~\cite{Nature.527.495.Soluyanov2015}, nearly flat bands, and saddle-point van Hove singularities (vHS).
These characteristics are closely related to topological and correlation physics. 
It is confirmed that the 2D BPN lattice constitutes a $\mathbb{Z}_{2}$ topological material, hosting topologically protected boundary states~\cite{doi:10.1021/acs.nanolett.2c00528.Son, PhysRevB.104.235422}.
Furthermore, the presence of saddle point vHS and nearly flat bands provide enhanced density of states (DOS), implying the potential emergence of correlation-related physics in the 2D BPN lattice.
In this context, the possibility of magnetic ordering in a pristine 2D BPN monolayer has been investigated using the density functional theory (DFT) calculations~\cite{doi:10.1021/acs.nanolett.2c00528.Son}. It is known that magnetic ordering can be induced through edge geometry modifications~\cite{doi:10.1021/ja710407t, Son2006, PhysRevLett.102.227205}, defect introduction~\cite{PhysRevLett.93.187202, PhysRevB.75.125408}, hydrogenation~\cite{doi:10.1021/nn4016289, doi:10.1021/nl9020733, 10.1063/1.3589970}, and adatom incorporation~\cite{PhysRevLett.91.017202} in carbon materials like graphene, but the presence of magnetism in pristine carbon materials is rare~\cite{GAO202111}. The 2D BPN lattice can be a promising candidate for magnetic all-carbon materials. 
Understanding magnetic ordering in BPN layers is crucial for their potential application as future spintronics platforms~\cite{doi:10.1021/jacs.2c02178}.

To understand correlation physics such as magnetism, it is of vital importance to consider many-body electron interactions in DFT calculations. To incorporate many-body electron correlations into DFT calculations, it is required to use advanced methods going beyond the mean-field level of the conventional correlation-exchange functionals such as the local density approximation (LDA) and the generalized gradient approximation (GGA).  
Correlation physics such as magnetic phase transitions can be investigated through the use of a newly developed DFT+$U$+$V$ method~\cite{Leiria_Campo_2010, PhysRevB.103.045141, PhysRevB.98.085127, PhysRevMaterials.5.104402, PhysRevResearch.2.043410, PhysRevB.102.155117}. 
The DFT+$U$+$V$ method includes not only on-site Hubbard interactions ($U$), but also inter-site interactions ($V$), thereby capturing the intricate interplay between strong electron localization and hybridization of extended neighboring orbitals, particularly for systems featuring covalent bonding characters. 
Remarkably, this extended Hubbard approach has demonstrated its capability to capture the effect of Coulomb interactions in electronic structure calculations at the similar level of $GW$ approximations~\cite{PhysRev.139.A796}, with applications spanning not only three-dimensional solids~\cite{Leiria_Campo_2010, PhysRevB.103.045141, PhysRevB.102.155117, PhysRevResearch.2.043410}, but also low-dimensional materials such as graphene~\cite{PhysRevB.102.155117} and black phosphorus~\cite{PhysRevResearch.2.043410}. 

When investigating possible magnetic ordering of BPN layers, one crucial consideration is accurate estimation of atomic structures. 
Influenced by atomic structure details, the positions of saddle-point vHS and nearly flat bands with respect to the Fermi energy $E_{F}$ are determined.  
To this end, accurate atomic structure predictions for BPN layers are imperative, considering electron correlations within the DFT+$U$+$V$ method. 
Regarding the potential impact of extended Hubbard corrections in introducing additional forces, it is possible that lattice constants and atomic configurations may deviate from those predicted by the mean-field level of DFT calculations. 
Noteworthy advancements have recently focused on forces corrected by extended Hubbard terms, enabling successfully calculations of lattice dynamics with extended Hubbard interactions~\cite{PhysRevB.104.104313.Wooil.Yang, JPCM34_295601_Yang_2022, PhysRevLett.130.136401}. 
The force calculation scheme, encompassing extended Hubbard forces, can be readily used to obtain fully relaxed atomic structures within the DFT+$U$+$V$ calculations. 

In this study, we conducted a comprehensive investigation into magnetic transitions of BPN layers by the DFT+$U$+$V$ method. 
We undertook the relaxation of atomic structures for BPN monolayer and bilayer, employing the extended Hubbard forces. 
By analyzing the positions of vHS and flat bands relative to $E_{F}$, we contrasted the electronic structures and magnetic ordering of BPN layers with and without the extended Hubbard forces. 
We showed that the subtle modification of the atomic structure due to the extended Hubbard forces predicts a non-magnetic state unlike the previous study~\cite{doi:10.1021/acs.nanolett.2c00528.Son}.
Subsequently, we explored their magnetic properties by subjecting them to external perturbations such as uniaxial strains and doping. 
\begin{figure}[t]
\begin{center}
\includegraphics[width=1.0\columnwidth, clip=true]{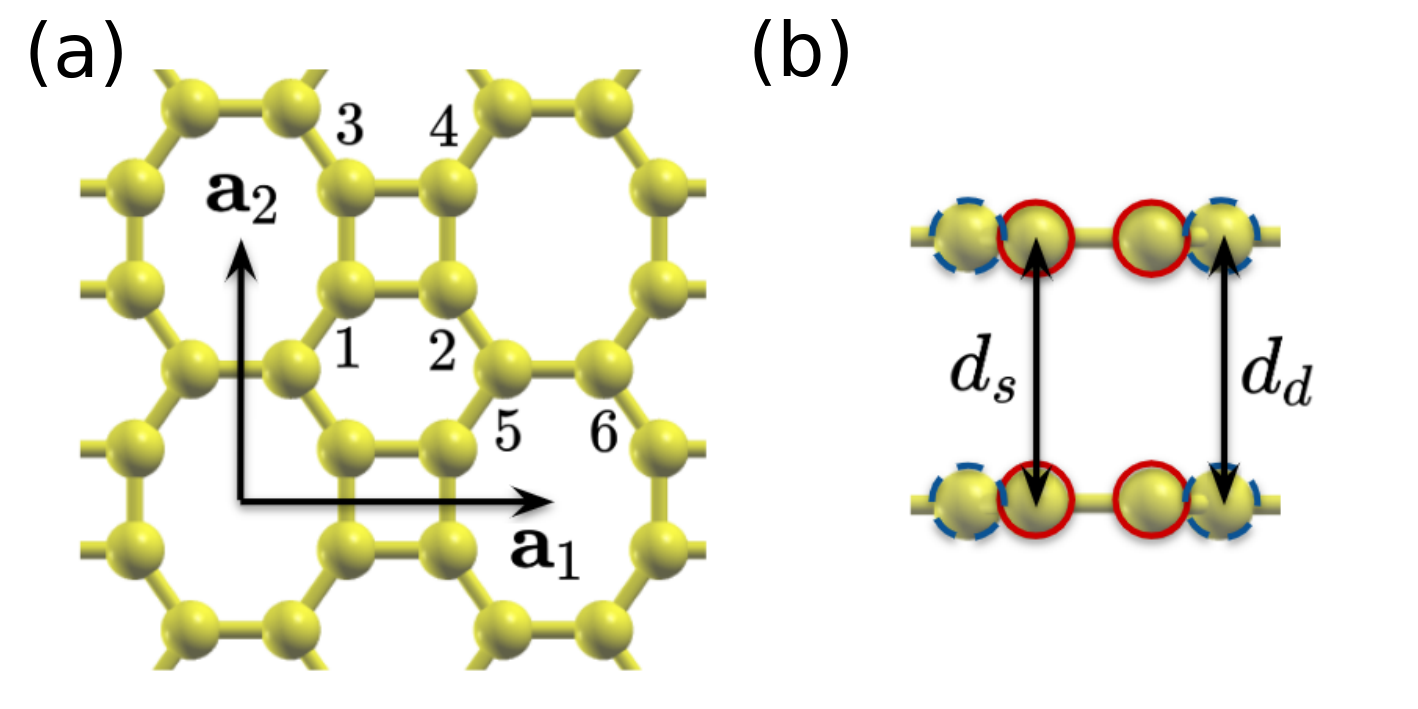} 
\end{center}
\caption{\label{fig:configuration} (Color online) (a) In-plane atomic configuration of BPN lattice. Carbon atoms are numbered from 1 to 6. Atoms from 1 to 4 belong to the square cluster, while atoms 5 and 6 constitute the dimer part. Lattice vectors $\mathbf{a}_{1}=a_{1} \hat{x}$ and $\mathbf{a}_{2}=a_{2} \hat{y}$ are indicated. (b) Side view of bilayer BPN. Atoms belonging to square cluster and dimer are surrounded by red solid lines and blue dashed ones. Inter-layer distances between square clusters (dimers) are denoted by $d_{s}$ ($d_{d}$). Atomic configurations are drawn by using XCrySDen~\cite{KOKALJ1999176}.}
\end{figure}

\section{\label{sec:DFT}Computational Details}
For DFT calculations we use \textsc{Quantum Espresso}~\cite{JPhys_CM_21_395502_2009,JPhys_CM_29_465901_2017} with the plane-wave (PW) basis, the PBE exchange-correlation functional~\cite{PhysRevLett_77_3865_1996} and norm-conserving pseudopotentials~\cite{PhysRevB_88_085117_2013, ComPhysComms_226_39_2018_Setten}. 
We also use a van der Waals (vdW) correction known as rev-vdW-DF2~\cite{PhysRevB.89.121103, PhysRevB.95.180101}.
We adopt $16\times16\times1$ $k$-point mesh, and the kinetic energy cutoff $100$ Ry for self-consistent calculations. To prevent interactions between periodic image layers, a vacuum region with a size of 15 $\textrm \AA$ is employed. 
To incorporate many-body electron correlations, we employed the DFT+$U$+$V$ method recently implemented in \textsc{Quantum Espresso}~\cite{PhysRevResearch.2.043410}, where on-site and inter-site Hubbard interactions are self-consistently determined by using the Hartree-Fock formalism based on the pseudohybrid Hubbard density functional known as ACBN0~\cite{PhysRevX.5.011006}. All atomic structures are relaxed to ensure that all components of forces acting on atoms are below 0.051 eV/{\AA}. Note that we call conventional DFT calculations with GGA functionals as DFT-GGA in order to distinguish against the DFT+$U$+$V$ method. 
\begin{figure}[t]
\begin{center}
\includegraphics[width=1.0\columnwidth, clip=true]{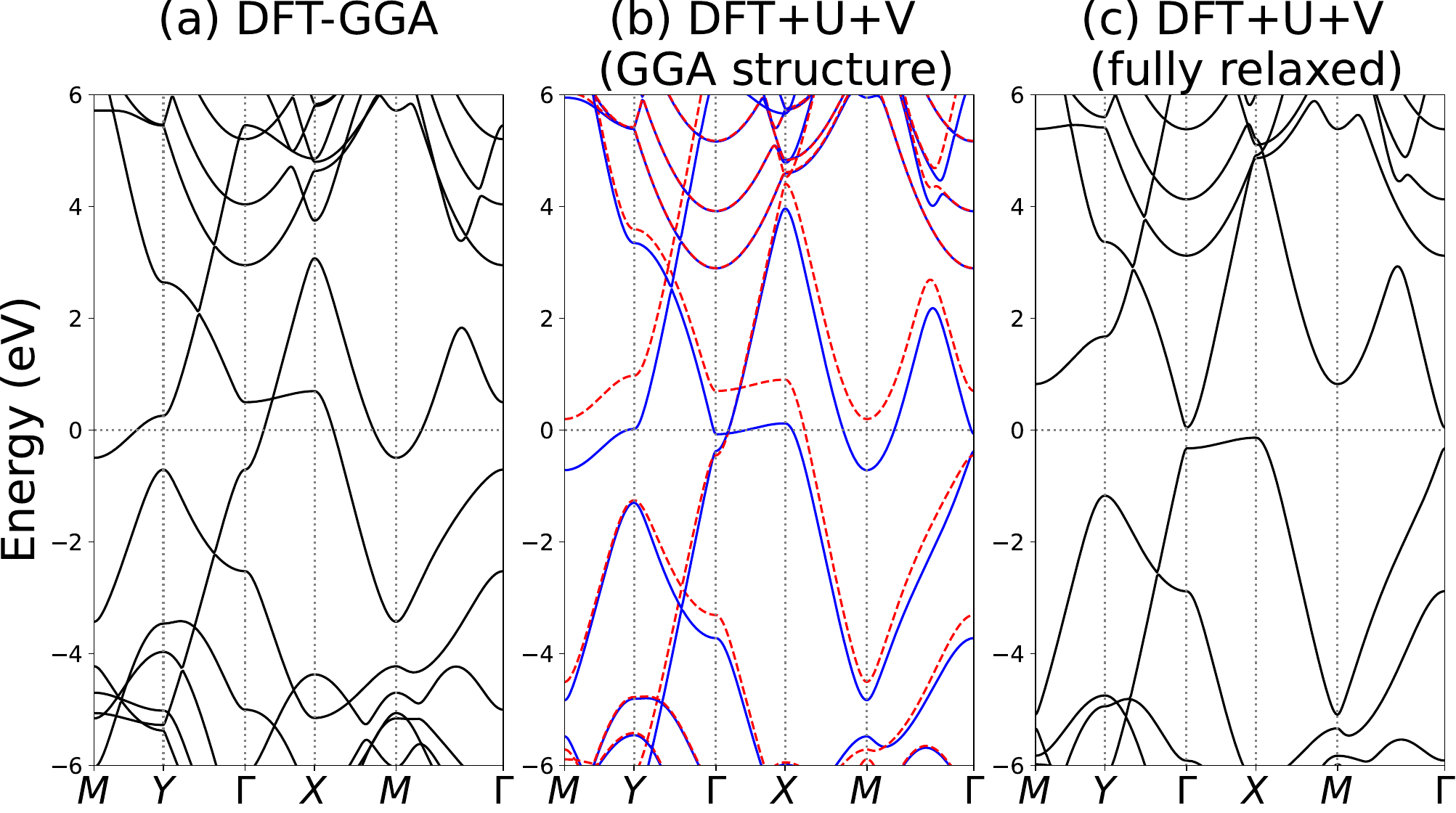} 
\end{center}
\caption{\label{fig:sBPN-bands} (Color online) Band structures of monolayer BPN (a) Band structures calculated by DFT-GGA. (b) Band structures computed from the DFT+$U$+$V$ method by using the GGA structure (c) Band structures of monolayer BPN fully relaxed by the DFT+$U$+$V$ method. Non-magnetic energy bands are indicated by black solid lines. Energy bands of the magnetic state are drawn by blue solid lines and red dashed ones, which correspond to opposite spin orientations. Here $E_{F}$ is set to be zero.}
\end{figure}

\section{Results and Discussion}
\begin{table*}[t]
\begin{ruledtabular}
\begin{tabular}{ c | c c c c c c c c c c c } 
monolayer & $|\mathbf{a}_{1}|$ & $|\mathbf{a}_{2}|$  & $d_{12}$ & $d_{13}$ & $d_{56}$ & $d_{25}$ & $\theta_{123}$ & $\theta_{234}$ & $\theta_{256}$ & & \\
\hline
DFT-GGA     & 4.52 & 3.77 & 1.46 &  1.46 & 1.45 & 1.41 & $90^{\circ}$ & $90^{\circ}$ & $124.9^{\circ}$ & & \\
DFT+$U$+$V$ & 4.47 & 3.80 & 1.41 &  1.51 & 1.46 & 1.40 & $90^{\circ}$ & $90^{\circ}$ & $124.8^{\circ}$ & & \\
\hline
\hline
bilayer & $|\mathbf{a}_{1}|$ & $|\mathbf{a}_{2}|$  & $d_{12}$ & $d_{13}$ & $d_{56}$ & $d_{25}$ & $\theta_{123}$ & $\theta_{234}$ & $\theta_{256}$ & $d_{s}$ & $d_{d}$ \\
\hline
DFT-GGA     & 4.52 & 3.77 & 1.46 &  1.46 & 1.45 & 1.41 & $90^{\circ}$ & $90^{\circ}$ & $124.9^{\circ}$ & 3.24 & 3.28 \\
DFT+$U$+$V$ & 4.47 & 3.81 & 1.41 &  1.51 & 1.47 & 1.40 & $90^{\circ}$ & $90^{\circ}$ & $124.8^{\circ}$ & 3.48 & 3.48 \\
\end{tabular}
\end{ruledtabular}
\caption{\label{table:config_parameters} Atomic structure parameters of monolayer and bilayer BPN optimized by DFT-GGA and DFT+$U$+$V$ calculations.}
\end{table*}
\subsection{\label{sec:Structures}Relaxed structures}
First we compare the atomic structures of both the BPN monolayer and bilayer, which are relaxed using DFT-GGA calculations as well as the DFT+$U$+$V$ method. 
Figure~\ref{fig:configuration} illustrates the atomic configurations of BPN layers.
The BPN lattice layer contains six carbon atoms in the unit cell. 
As depicted in Fig.~\ref{fig:configuration}(a), the carbon atoms are labeled accordingly, with atoms 1 to 4 forming a square cluster, and atoms 5 and 6 forming a dimer. 
For the monolayer BPN, all six atoms reside within the same plane, without any out-of-plane displacement. 
The BPN lattice exhibits a pattern consisting of square, hexagon, and octagon rings. 
Additionally, when stacking BPN layers, each layer precisely aligns on top of its neighboring layer, as demonstrated in Fig.~\ref{fig:configuration}(b). 

Table~\ref{table:config_parameters} presents a comparison of lattice constants and atomic configurations of BPN lattices, which are relaxed using GGA and DFT+$U$+$V$ methods. 
With the inclusion of $U$+$V$ corrections, the lattice constant $a_{1}$ decreases from 4.52 {\AA} to 4.47 {\AA}, while the lattice constant $a_{2}$ increases from 3.77 {\AA} to 3.80 {\AA}. 
Consequently, the DFT+$U$+$V$ method predicts that a square with DFT-GGA is transformed to the rectangle cluster, experiencing contraction along the $x$-axis and elongation along the $y$-axis. 
Furthermore, the size of the dimer slightly increases by 0.02 {\AA}. 

For the BPN bilayer, the energy minimum configuration is the on-top stacking where the upper layer is exactly on the top of the lower layer as shown in Fig.~\ref{fig:configuration}. 
both DFT-GGA and DFT+$U$+$V$ calculations indicate the intra-layer configurations remain the same as those of the monolayer. 
However, the inclusion of $U$+$V$ corrections leads to changes in inter-layer configurations. 
In the DFT-GGA calculations, carbon atoms in the bilayer exhibit small out-of-plane displacements. 
The inter-layer distances between atoms in the square cluster and the dimer, denoted as $d_{s}$ and $d_{d}$ respectively, are found to be $d_{s}=3.24$ {\AA} and $d_{d}=3.28$ {\AA} as indicated in Fig.~\ref{fig:configuration}(b) and Table~\ref{table:config_parameters}.
When $U$+$V$ corrections are incorporated, the inter-layer distance undergoes an increase of approximately 0.24 {\AA}, and out-of-place displacements disappear.

\begin{figure}[t]
\begin{center}
\includegraphics[width=1.0\columnwidth, clip=true]{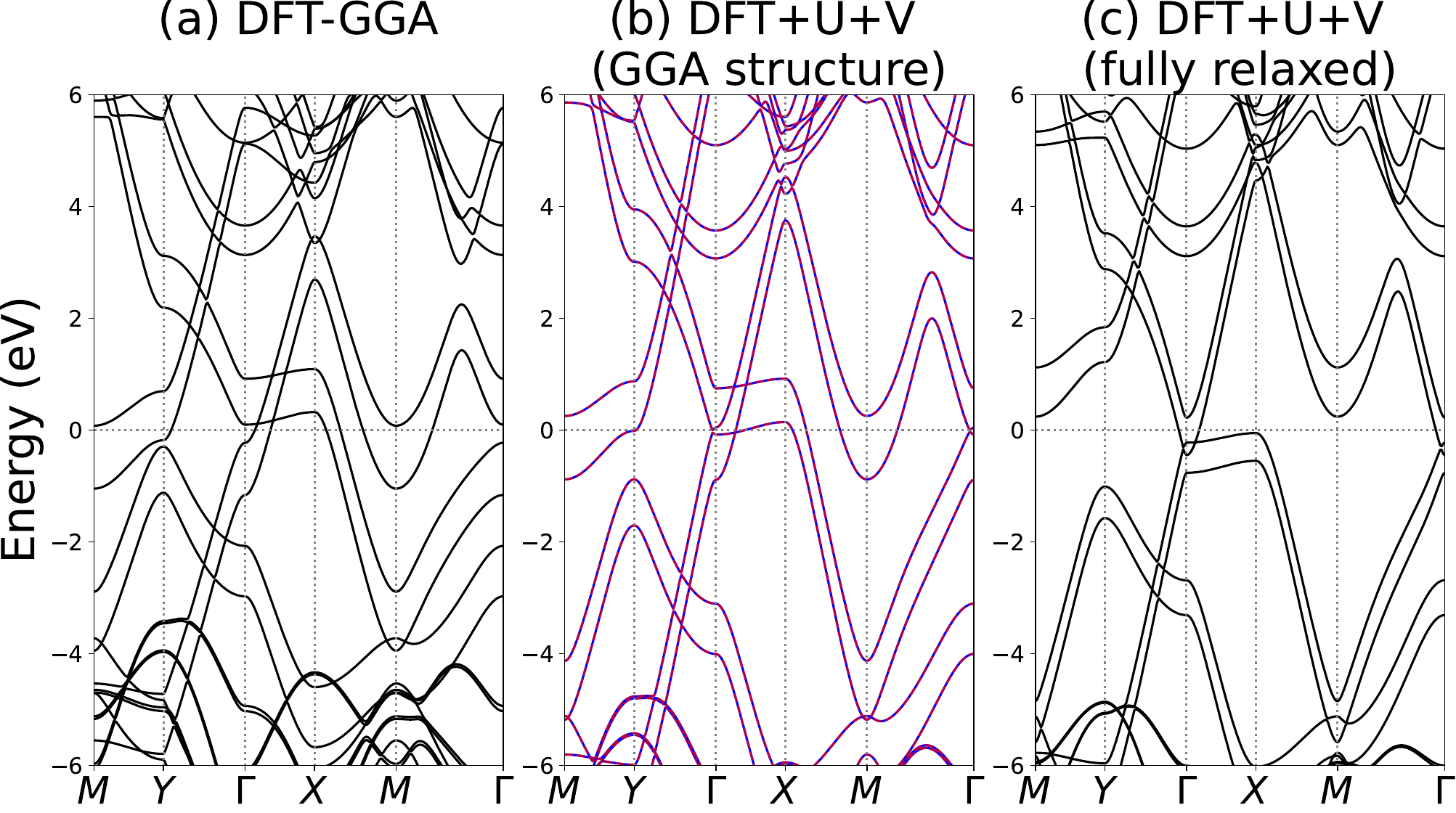} 
\end{center}
\caption{\label{fig:dBPN-bands} (Color online) Band structures of bilayer BPN (a) Band structures calculated by DFT-GGA. (b) Band structures computed from the DFT+$U$+$V$ method by using the GGA structure (c) Band structures of monolayer BPN fully relaxed by the DFT+$U$+$V$ method. Non-magnetic energy bands are indicated by black solid lines. Energy bands of the magnetic state are drawn by blue solid lines and red dashed ones, which correspond to opposite spin orientations.}
\end{figure}
\subsection{\label{sec:Band_Sturcture}Band Structures}
We observed distinct electronic structure properties of monolayer BPN, predicted by DFT-GGA and DFT+$U$+$V$ calculations, as demonstrated in Fig.~\ref{fig:sBPN-bands}.  
We reproduced the DFT-GGA electronic band structures as reported Ref.~\cite{doi:10.1021/acs.nanolett.2c00528.Son, PhysRevB.105.035408}. 
DFT-GGA calculations reveal a saddle-shaped band hosting the van Hove singularity (vHS) at $\Gamma$, intersecting with the nearly flat band around $\frac{2}{5}\Gamma X$ in Fig.~\ref{fig:sBPN-bands}(a). 
The band structures in the vicinity of the crossing point exhibit a type-II Dirac state~\cite{Nature.527.495.Soluyanov2015, Nat.Commun.8.257.2017.Yan, RevModPhys.90.015001.2018,PhysRevB.96.041201,PhysRevLett.119.016401}. 
The nearly flat band 
is mainly composed of $p_{z}$ orbitals localized on the square cluster. The saddle-shaped band is formed by the combination of $p_{z}$ orbitals on the square unit and the dimer, with the main contribution coming from $p_{z}$ orbitals of the dimer.
An additional remarkable feature of the band structure is the presence of a saddle-shaped band that hosts another vHS near the $Y$ point around $E_{F}$.

Next we perform DFT+$U$+$V$ calculations using the atomic structure relaxed by DFT-GGA, in order to differentiate them from the full DFT+$U$+$V$ calculations where atomic structures are entirely relaxed using $U$+$V$ corrections. 
Here we call the atomic structures relaxed by DFT-GGA as GGA structures. The self-consistent calculations of onsite ($U$) and inter-site ($V$) Hubbard interactions are $U_{t}=5.96$ eV, $U_{d}=6.01$ eV, $V_{12}=V_{34}=V_{13}=V_{24}=3.06$ eV, $V_{56}=3.11$ eV, and $V_{25}=3.14$ eV, where $t$ and $d$ denote the tetragon and the dimer respectively and atomic site numbers in inter-site interactions follow Fig.~\ref{fig:configuration}(a). 
In comparison with DFT-GGA calculations, the inclusion of $U$+$V$ corrections leads to increased overall bandwidths of the resulting band structures. 
The DFT+$U$+$V$ calculations with GGA structures reveal a ferrimagnetic ground state for the single-layer BPN, where the square unit and the dimer possess opposite magnetic moments. 
For example, Fig.~\ref{fig:rho}(a) represents spin polarization density $\Delta \rho \equiv \rho_{\uparrow} - \rho_{\downarrow}$ of the ferrimagnetic phase, where $\rho_{\uparrow}$ and $\rho_{\downarrow}$ are spin-up and spin-down densities, respectively. 
Since their magnetic moments do not completely cancel out, monolayer BPN exhibits a net magnetic moment.
Figure~\ref{fig:sBPN-bands}(b) illustrates the spin-resolved band structures of the ferrimagnetic ground state, displaying opposite spin orientations. 
In comparison to DFT-GGA band structures, vHS moves closer to $E_{F}$, and one nearly flat band at $\Gamma X$ shifts downwards towards $E_{F}$. 
Accordingly, among four crossing points between two saddle-shaped bands and two nearly flat bands, two of them are located in the vicinity of $\Gamma$-point and have energies close to $E_{F}$. 

However, when the atomic structure is fully relaxed with the DFT+$U$+$V$ method, the corresponding band structures exhibit different electronic properties compared to DFT+$U$+$V$ calculations with GGA structures. 
Firstly, the resulting ground state is non-magnetic, which is in contrast to the fact that the DFT+$U$+$V$ calculations with GGA structures predict a ferrimagnetic ordering.
Secondly, as illustrated in Fig.~\ref{fig:sBPN-bands}(c), a band gap opens at the zone center even in the fully relaxed structure without external perturbations such as strains and doping. 
It is found that the band edge minimum at $M$ moves upward above $E_{F}$, resulting in the single layer BPN being a semiconductor with an indirect band gap between the conduction band bottom at $\Gamma$ and the valence band top at $X$. 
We also note that the self-consistent Hubbard interactions for the full DFT+$U$+$V$ calculation are $U_{t}=5.99$ eV, $U_{d}=6.00$ eV, $V_{12}=V_{34}=3.14$ eV, $V_{13}=V_{24}=3.00$ eV, $V_{56}=3.07$ eV, and $V_{25}=3.15$ eV. 

\begin{figure*}[t]
\begin{center}
\includegraphics[width=1.0\textwidth, clip=true]{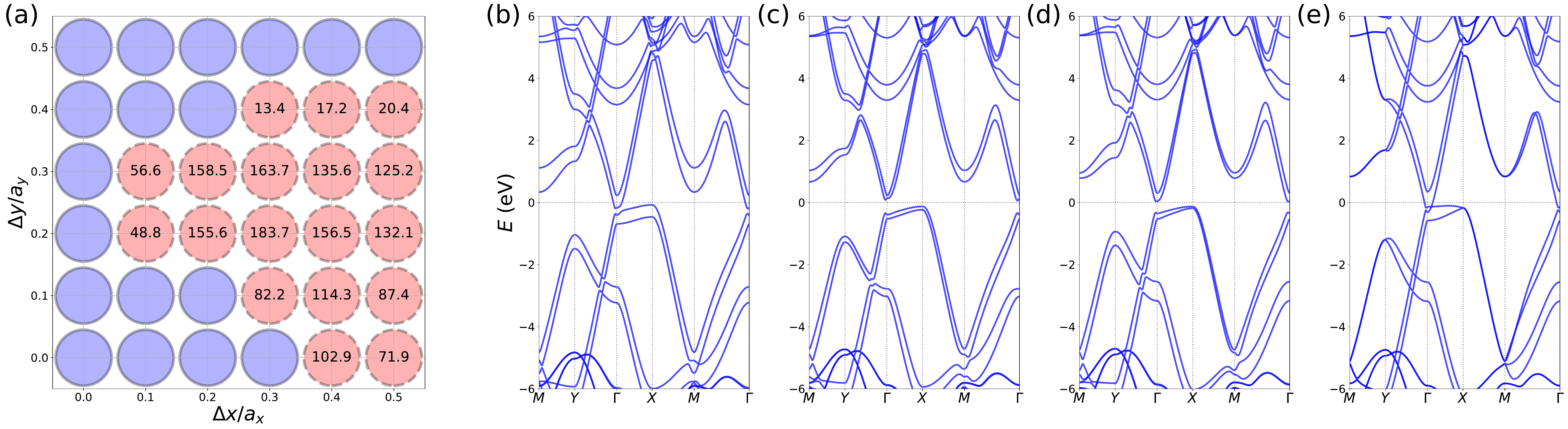} 
\end{center}
\caption{\label{fig:dBPN-slide} (Color online) Electronic structures of stacking configurations for bilayer BPNs: (a) Band gaps depending on $\Delta x$ and $\Delta y$. Blue circles with solid edge lines (red circles with dashed edge lines) represent that the corresponding stacking configuration is metal (semiconductor). For semiconducting configurations, band gap energy is written inside the circle in a unit of meV. Band structures for (a) $\Delta x/a_{x}$ = 0.1 and $\Delta y/a_{y}$ = 0.1, (b) $\Delta x/a_{x}$ = 0.3 and $\Delta y/a_{y}$ = 0.2, (c) $\Delta x/a_{x}$ = 0.4 and $\Delta y/a_{y}$ = 0.2, and (d) $\Delta x/a_{x}$ = 0.5 and $\Delta y/a_{y}$ = 0.5. Here $E_{F}$ is set to be zero.}
\end{figure*}

We extend our discussions to electronic band structures of bilayer BPN, comparing DFT-GGA calculations and DFT+$U$+$V$ ones. 
In the DFT-GGA calculations, bilayer BPN is found to be non-magnetic. 
The resulting band structures, as shown in Fig.~\ref{fig:dBPN-bands}(a), exhibit two nearly flat bands along $\Gamma X$ direction and two bands hosting vHS at the $\Gamma$ point. 
These bands intersect at four points along $\Gamma X$. 

Using atomic structures relaxed by DFT-GGA (GGA structures), DFT+$U$+$V$ calculations reveal that bilayer BPN exhibits an antiferromagnetic ground state. 
Each layer displays a ferrimagnetic order similar to monolayer BPN in the same calculation condition, but the upper and lower layers possess opposite spin orientations. 
The self-consistent evaluations of Hubbard interactions are $U_{t}=5.98$ eV, $U_{d}=6.02$ eV, $V_{12}=V_{34}=V_{13}=V_{24}=3.06$ eV, $V_{56}=3.11$ eV, and $V_{25}=3.14$ eV, which are almost the same with those of monolayer BPN in DFT+$U$+$V$ calculations with GGA structure. 
The band structures from DFT+$U$+$V$ calculations using the GGA structures, as depicted in Fig.~\ref{fig:dBPN-bands}(b), show each band to be doubly degenerate with opposite spin configurations. 
The crossing point between the upper band with vHS and the lower flat band is annihilated with its time-reversal partner at $\Gamma$, so there are three type-II Dirac points.  

Fully relaxed the atomic configuration with the DFT+$U$+$V$ method, 
the bilayer BPN reverts to a non-magnetic state. 
In Fig.~\ref{fig:dBPN-bands}(c), all type-II Dirac points merge with their time-reversal partners, leaving no type-II Dirac point remaining. 
Around $E_{F}$, there are two V-shaped bands on the conduction side and two trapezoid-shaped valence bands, which include flat bands on $\Gamma X$.  
Unlike monolayer BPN, the bilayer BPN is metallic due to overlap between lower V-shaped band and upper flat band around $E_{F}$. 
Note that the self-consistent Hubbard interactions for the full DFT+$U$+$V$ calculation for bilayer BPN are the same with those of monolayer BPN.

We also consider electronic structures of other stacking configurations as well as the energy-minimum on-top configuration. We shift the $x$ and $y$ positions of the upper layer relative to the lower layer by $\Delta x$ and $\Delta y$.   
Depending on $\Delta x$ and $\Delta y$, it is found that bilayer BPN can be either metal or semiconductor, whose band gap is about from 10 meV to 200 meV as shown in Fig.~\ref{fig:dBPN-slide}(a). 
Regardless of stacking configurations, bilayer BPN remains non-magnetic.

\begin{figure}[t]
\begin{center}
\includegraphics[width=1.0\columnwidth, clip=true]{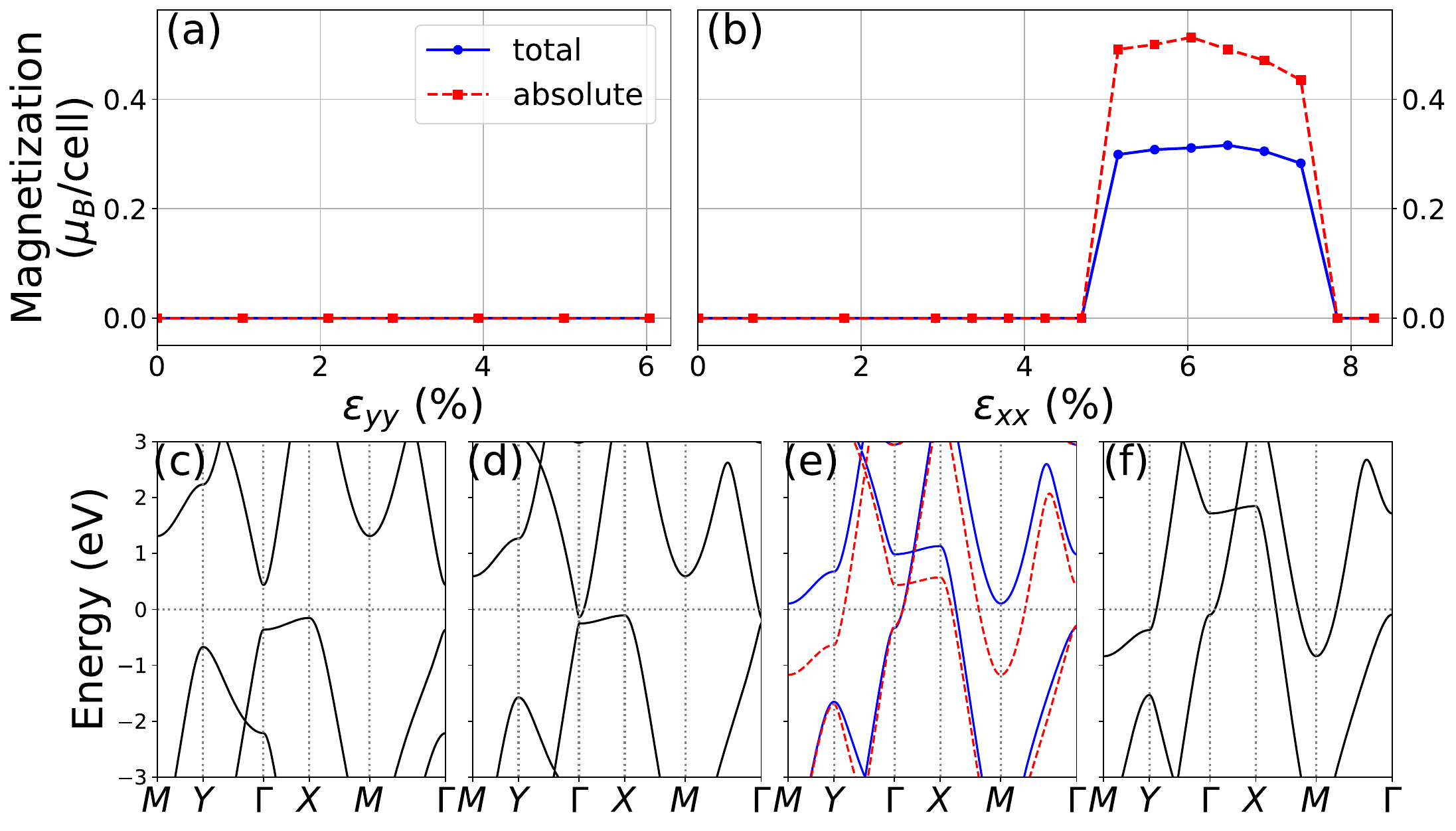} 
\end{center}
\caption{\label{fig:strained_sBPN_mag} (Color online) Magnetization of single-layer BPN as a function of uniaxial strains (a) $\varepsilon_{yy}$ and (b) $\varepsilon_{xx}$. Absolute and total magnetizations are denoted by red dashed lines with square markers and blue solid lines with closed circle markers, respectively. Band structures with uniaxial strains (c) $\varepsilon_{yy}=2.89$ ($a_{2}=3.92$) (d) $\varepsilon_{xx}=3.80$ ($a_{1}=4.64$) (e) $\varepsilon_{xx}=6.04$ ($a_{1}=4.74$) (f) $\varepsilon_{xx}=8.28$ ($a_{1}=4.84$). Strains and lattice constants are in unit of \% and \AA, respectively. Black solid lines indicate non-magnetic energy bands. Energy bands of the magnetic state are drawn by blue solid lines and red dashed ones, which correspond to opposite spin orientations. Here $E_{F}$ is set to be zero.}
\end{figure}

\begin{figure}[t]
\begin{center}
\includegraphics[width=1.0\columnwidth, clip=true]{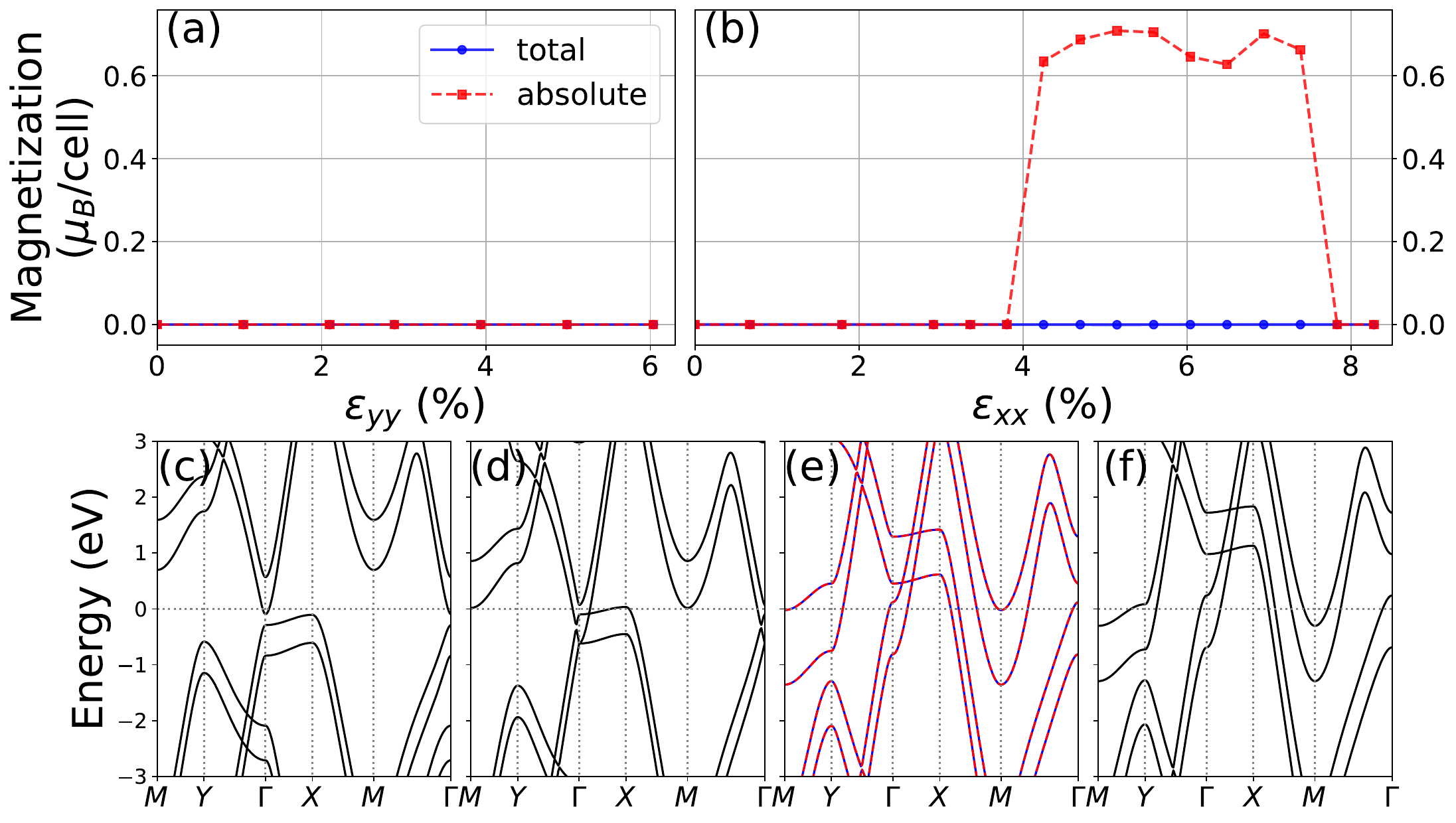} 
\end{center}
\caption{\label{fig:strained_dBPN_mag} (Color online) Magnetization of bilayer BPN as a function of uniaxial strains (a) $\varepsilon_{yy}$ and (b) $\varepsilon_{xx}$. Absolute and total magnetizations are denoted by red dashed lines with square markers and blue solid lines with closed circle markers, respectively. Band structures with uniaxial strains (c) $\varepsilon_{yy}=2.89$ ($a_{2}=3.92$) (d) $\varepsilon_{xx}=3.36$ ($a_{1}=4.62$) (e) $\varepsilon_{xx}=6.04$ ($a_{1}=4.74$) (f) $\varepsilon_{xx}=7.83$ ($a_{1}=4.82$). Strains and lattice constants are in unit of \% and \AA, respectively. Black solid lines indicate non-magnetic energy bands. Energy bands of the magnetic state are drawn by blue solid lines and red dashed ones, which correspond to opposite spin orientations. Here $E_{F}$ is set to be zero.}
\end{figure}

\begin{figure}[t]
\begin{center}
\includegraphics[width=1.0\columnwidth, clip=true]{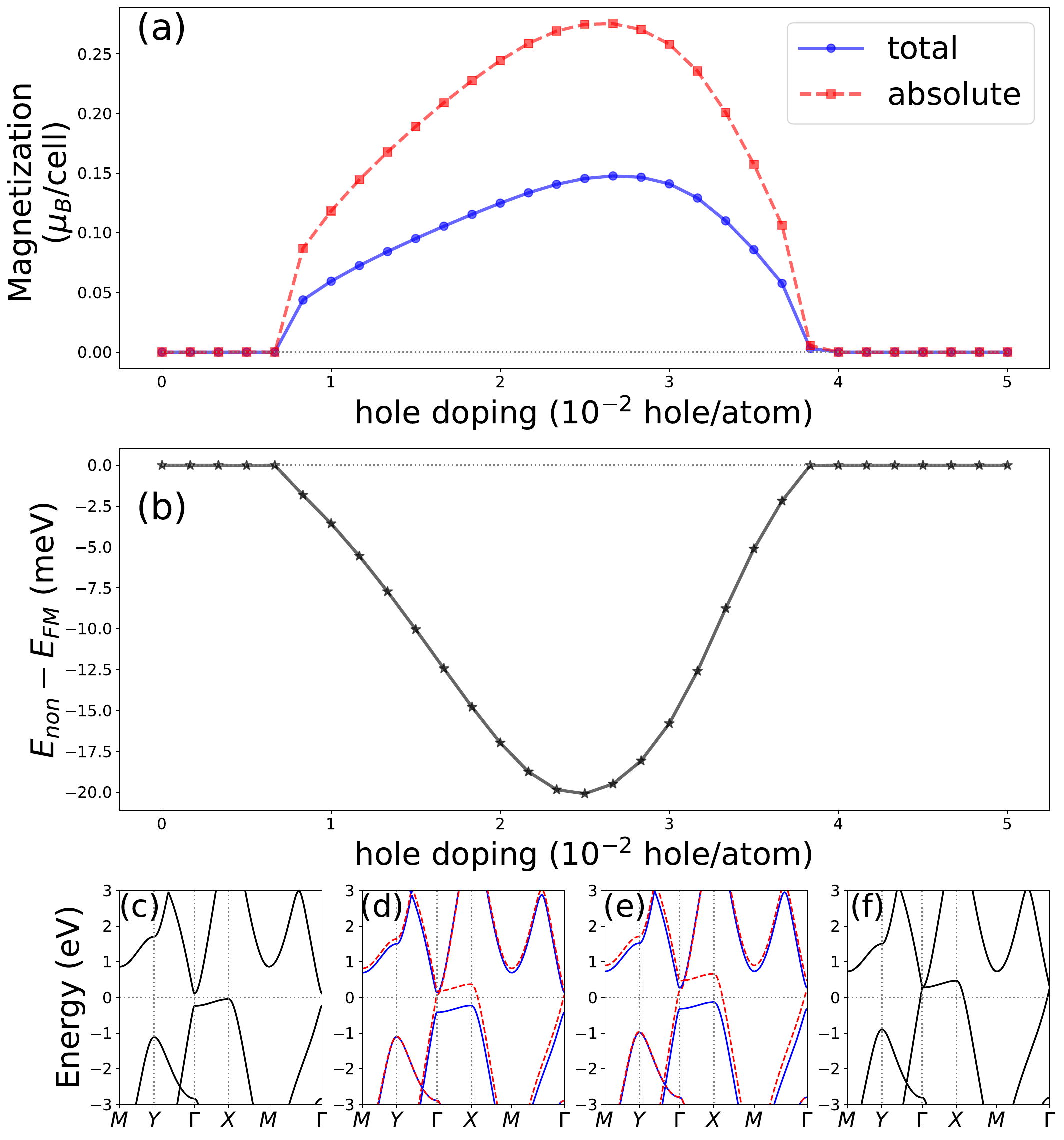} 
\end{center}
\caption{\label{fig:doped-sBPN} (Color online) (a) Magnetization of single-layer BPN as a function of hole doping concentration. Red dashed lines with square markers (blue solid lines with close circles) denote absolute (total) magnetization. (b) Total energy differences between non-magnetic and magnetic phases. Band structures at hole doping concentrations (c) 0.33 (d) 1.67 (e) 2.50 (f) 4.17 in unit of $10^{-2}$ hole per atom. Black solid lines indicate non-magnetic energy bands. Energy bands of the magnetic state are drawn by blue solid lines and red dashed ones, which correspond to opposite spin orientations. Here $E_{F}$ is set to be zero.}
\end{figure}
\begin{figure}[t!]
\begin{center}
\includegraphics[width=1.0\columnwidth, clip=true]{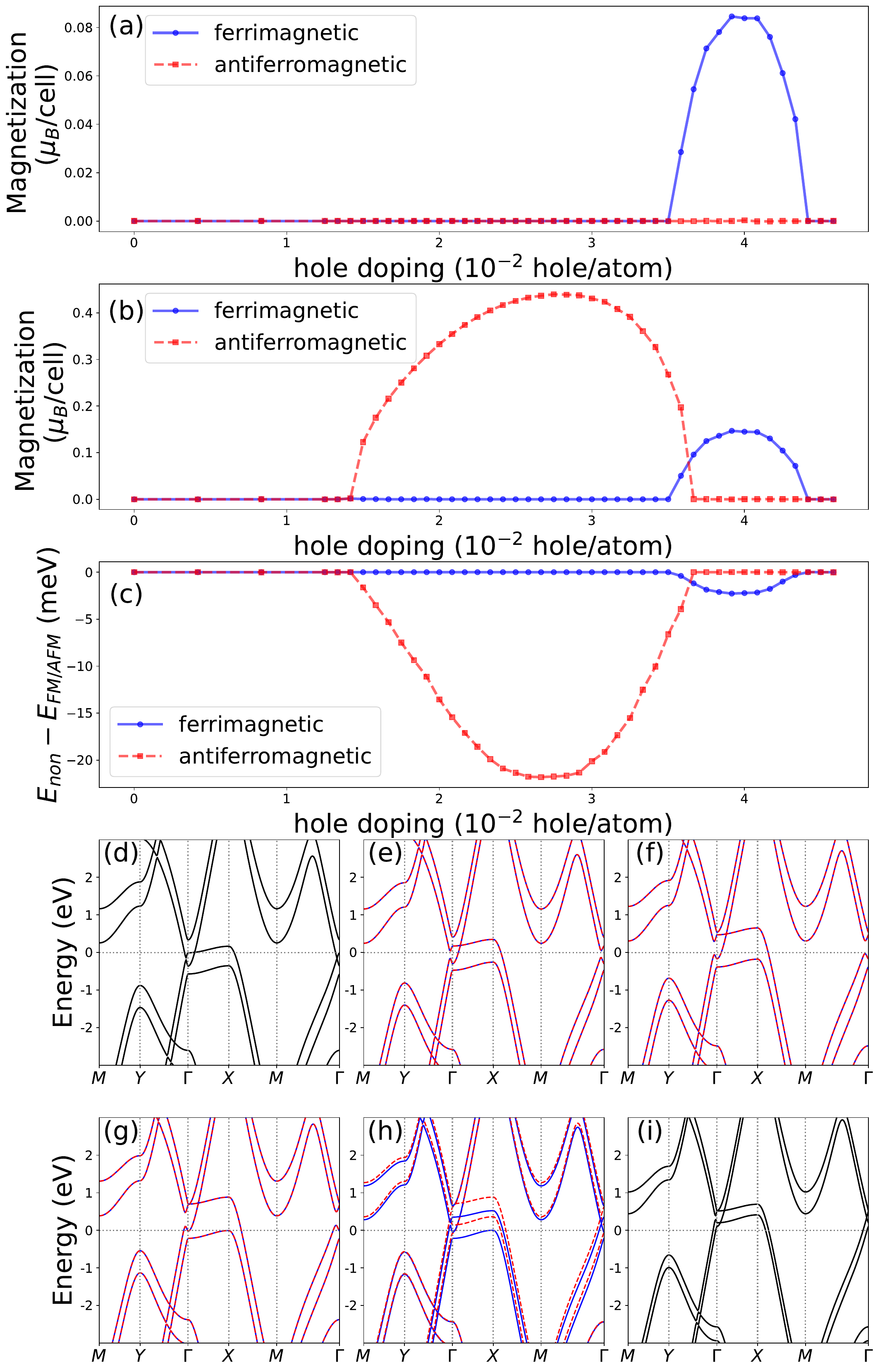} 
\end{center}
\caption{\label{fig:doped-dBPN} (Color online) (a) Total magnetization of bilayer BPN as a function of hole doping concentration. (a) Absolute magnetization of bilayer BPN as a function of hole doping concentration. (c) Total energy differences between non-magnetic and magnetic phases. Red dashed lines with square markers (blue solid lines with close circles) denote the total magnetization of the antiferromagnetic state (ferrimagnetic state) for (a), (b), and (c). Band structures of bilayer BPN at hole doping concentrations (d) 0.83 (e) 1.67 (f) 2.50 (g) 3.33 (h) 3.75 (i) 4.58 in a unit of $10^{-2}$ hole/atom. Black solid lines indicate non-magnetic energy bands. Energy bands of the magnetic state are drawn by blue solid lines and red dashed ones, which correspond to opposite spin orientations. Here $E_{F}$ is set to be zero.}
\end{figure}
\subsection{Magnetic transitions}
It is shown that monolayer and bilayer BPN fully relaxed within the DFT+$U$+$V$ method are non-magnetic, which is in contrast to the fact that monolayer and bilayer BPN exhibit magnetic ordering when the DFT+$U$+$V$ method is employed with GGA structures. 
The key distinction arises from the relative positions of intriguing electronic structures such as nearly flat bands and saddle-point vHS with respect to $E_{F}$. 
This observation suggests a strategy to induce magnetic ordering in BPN layers: Tuning positions of nearly flat bands and saddle-point vHS, which can be achieved through external perturbations like uniaxial strains and hole doping. 

\subsubsection{Uniaxial strains}
We investigate the impact of uniaxial strains on the magnetic transitions of monolayer and bilayer BPN. 
External strains cause shifts in the positions of vHS and flat bands relative to $E_{F}$, potentially leading to magnetic transitions. 
We observed that the GGA structure, which possesses a larger lattice constant $a_{1}$ than the DFT+$U$+$V$ structure, exhibits magnetism with the $U$+$V$ correction. 
This observation suggests that enlarging the unit cell may induce a magnetic transition. 

Figures~\ref{fig:strained_sBPN_mag}(a) and (b) present total and absolute magnetization as a function of $\varepsilon_{yy}$ and $\varepsilon_{xx}$, respectively. 
Here the uniaxial strain $\varepsilon_{xx}$ ($\varepsilon_{yy}$) along $x$ ($y$) direction is expressed as
$(a_{1}-a_{1}^{0})/a_{1}^{0}$ ($(a_{2}-a_{2}^{0})/a_{2}^{0}$), where $a_{1}^{0}$ and $a_{2}^{0}$ represent the lattice constants along $x$ and $y$ directions without external strains, respectively, as summarized in Table.~\ref{table:config_parameters}.
When uniaxial $y$-strains are applied, we find that they do not induce a magnetic transition.
For example, as shown in Fig.~\ref{fig:strained_sBPN_mag}(c), the band structures at $\varepsilon_{yy}=2.89$ \% demonstrate that uniaxial $y$ strains lead to an enhancement of the band gap opening around $E_{F}$, resulting in a lack of DOS involved in the magnetic transition. 

In contrast, uniaxial strains along $x$ direction can trigger a transition from a non-magnetic state to a ferrimagnetic one.
The magnetic transition to a ferrimagnetic phase is observed at $\varepsilon_{xx} \approx 4.90$ \%.
For smaller $\varepsilon_{xx}$, type-II Dirac points are annihilated with vHS, leading to the opening of an energy gap at $\Gamma$. For example, at $\varepsilon_{xx}=3.80$ in Fig.~\ref{fig:strained_sBPN_mag}(d), a small energy gap is observed at $\Gamma$. 
However, as $\varepsilon_{xx}$ increases above 4.90 \%, Dirac points and vHS reappear around $E_{F}$, causing the energy gap at $\Gamma$ to close. 
At the same time, vHS at $Y$ and the conduction band edge at $M$ move downward to $E_{F}$.

The ferrimagnetic phase, characterized by total magnetization being smaller than absolute one, persists up to $\varepsilon_{xx} \approx 7.60$ \% as shown in Fig.~\ref{fig:strained_sBPN_mag}(b). For the ferrimagnetic phase, energy bands associated with opposite spin orientations are lifted, for example, as illustrated in Fig~\ref{fig:strained_sBPN_mag}(b) for $\varepsilon_{xx} \approx 6.04$ \%.
Beyond the strain level $\varepsilon_{xx} \approx 7.60$ \%, the single-layer BPN returns to a non-magnetic state. 
At $x$ strains higher than 7.60 \%, the two vHS points at $\Gamma$ and $Y$ are located near $E_{F}$, while the flat band on $\Gamma X$ moves further away from $E_{F}$. For example, see Fig.~\ref{fig:strained_sBPN_mag}(f).

Under uniaxial strains, bilayer BPN undergoes magnetic transitions similar to the single-layer case. 
uniaxial $y$ strains eliminate the small overlap between the lower V-shaped band and the upper flat band present in the fully relaxed structure, eventually leading to a semiconducting state with an indirect band gap. See Fig.~\ref{fig:strained_dBPN_mag}(c).   
No magnetic transition is observed as the DOS vanishes around $E_{F}$.  

Under the application of uniaxial $x$ strains, the bilayer BPN undergoes an antiferromagnetic transition, where the upper and lower layers have opposite spin orientations, resulting in the absence of total magnetization, for example, as shown in Fig.~\ref{fig:rho}(b).  
The magnetic transition takes place at $\varepsilon_{xx} \approx 4.0$ \%, which is smaller than the transition strain observed for the monolayer BPN.
When $\varepsilon_{xx} < 4.0$ \%, at which bilayer BPN is non-magnetic, some of type-II Dirac points and vHS around $\Gamma$ are partly restored, as illustrated in Fig.~\ref{fig:strained_dBPN_mag}(d) where  $\varepsilon_{xx} < 3.36$ \%.
Around the transition strain $\varepsilon_{xx} \approx 4.0$, all four crossing points between two flat bands and two saddle-shaped bands are fully recovered. 
For $\varepsilon_{xx} > 4.00$ \%, the bilayer BPN becomes antiferromagnetic, and this antiferromagnetic ordering persists up to $\varepsilon_{xx} \approx 7.6$. 
We also note that the two flat bands on $\Gamma X$ shift upward, and the conduction bands around $M$ and $Y$ move down, as $\varepsilon_{xx}$ increases. 

Note that we also consider other mechanical perturbations such as biaxial strains and in-plane and out-of-plane hydrostatic pressure. 
Applying biaxial strains up to 10 \% and in-plane hydrostatic pressure (up to 10\% lattice shrinkage), we do not observe magnetic transitions.
Using out-of-plane hydrostatic pressure on bilayer BPN, we find no magnetic transition with interlayer distance reduction up to 1.6 {\AA}. 

\subsubsection{Hole doping}
Considering that flat bands on $\Gamma X$ are below $E_{F}$ [Figs.~\ref{fig:sBPN-bands}(c) and \ref{fig:dBPN-bands}(c)], hole doping might induce magnetic transitions by aligning the flat bands around $E_{F}$.
To investigate this scenario, we performed DFT+$U$+$V$ calculations with different hole doping concentrations in single-layer and bilayer BPN, as shown in Figs.~\ref{fig:doped-sBPN} and \ref{fig:doped-dBPN}. 

Figure~\ref{fig:doped-sBPN}(a) presents that single-layer BPN becomes ferrimagnetic at a hole doping concentration $n_{h} \approx 7\times 10^{-3}$ hole per atom. In the ferrimagnetic phase, spins of rectangle clusters and dimers are oriented in opposite directions, similar to the ferrimagnetic ordering induced by uniaxial strains. 
The ferrimagnetic ordering is maintained up to about $n_{h} \approx 3.8\times 10^{-2}$ hole per atom. 

The band structures near $E_{F}$ show different evolutions for hole doping, compared to uniaxial strains. 
In the case of uniaxial strains, the ferrimagnetic phase transition is typically accompanied by the full restoration of type-II Dirac points and vHS around $\Gamma$, as shown in the previous section. 
However, the ferrimagnetic state induced by hole doping has different band structures around $E_{F}$ depending on spin orientations. 
For example, as shown in Fig.~\ref{fig:doped-sBPN}(d), the energy bands associated with one spin direction exhibit a saddle point at $\Gamma$ and Dirac points with the flat band on $\Gamma X$ located above $E_{F}$. 
Conversely, no Dirac point or vHS are recovered in the band structures with the opposite spin orientation. 
Additionally, it is worth noting that overall band structures, including the other vHS at $Y$ and conduction band edges at $M$, move upwards relative to $E_{F}$ as hole doping concentrations increase. 
This is in contrast to the case of uniaxial strains, where band structures around $YM$ evolve in the opposite direction compared with those on $\Gamma X$.

Hole doping also induces magnetic transitions in bilayer BPN, as illustrated in Fig.~\ref{fig:doped-dBPN}. 
The critical hole doping concentration for the antiferromagnetic transition is $n_{h} \approx 1.42 \times 10^{-2}$ hole per atom, which is larger than that of single-layer BPN. 
An interesting feature of hole-doped bilayer BPN is another magnetic phase transition from antiferromagnetic ordering to ferrimagnetic one at a higher hole doping concentration $n_{h} \approx 3.63 \times 10^{-2}$ hole per atom. 
In the ferrimagnetic state of bilayer BPN, each layer has ferrimagnetic ordering, where spins of rectangle clusters are aligned oppositely to those of dimers, which is the same as ferrimagnetic single-layer BPN. 
However, when considering spins at the same positions in the upper and lower layers, they are oriented in the same direction, as shown in Fig.~\ref{fig:rho}(c).  

Similar to hole-doped single-layer BPN, the magnetic transitions in hole-doped bilayer BPN are not associated with the full recovery of type-II Dirac points and saddle-shaped bands around $\Gamma$, as shown in Figs.~\ref{fig:doped-dBPN}(d)-(i). 
The type-II Dirac point is formed on the upper flat band on $\Gamma X$, while the lower flat band does not host a type-II Dirac point. 
For the ferrimagnetic state, energy bands associated with opposite spins are split, resulting in four flat bands on $\Gamma X$ in Fig.~\ref{fig:doped-dBPN}(h). 

\subsubsection{Magnetic interactions}
\begin{figure*}[t]
\begin{center}
\includegraphics[width=1.0\textwidth, clip=true]{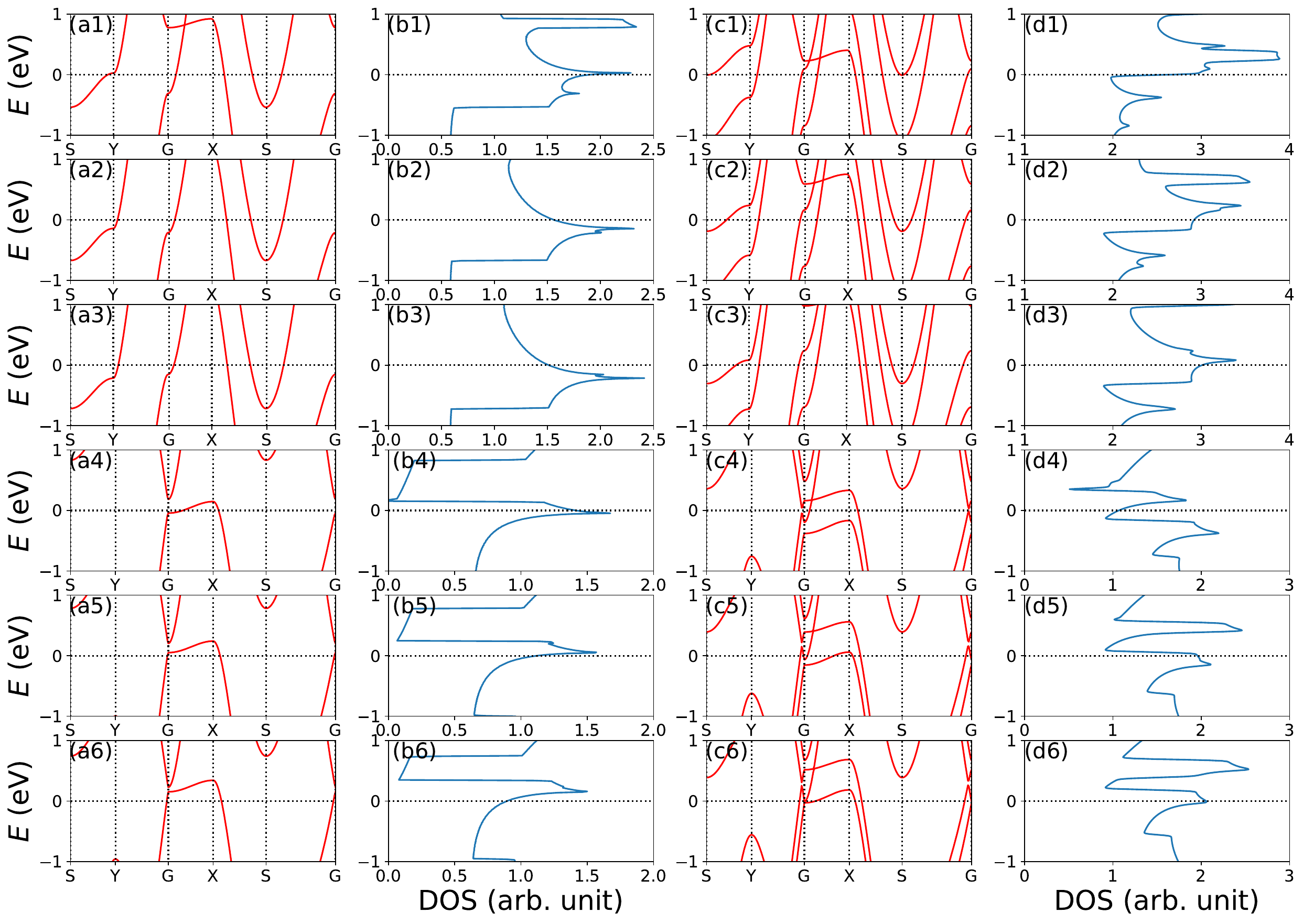} 
\end{center}
\caption{\label{fig:dos} (Color online) Band structures (the first column, blue solid lines) and corresponding DOS (the second column, red solid lines) for monolayer BPN with uniaxial strains and hole doping concentrations: (a1) and (b1) $\varepsilon_{xx}=5.6$ \%, (a2) and (b2) $\varepsilon_{xx}=6.5$ \%, (a3) and (b3) $\varepsilon_{xx}=7.4$ \%, (a4) and (b4) $n_{h}=1.7\times 10^{-2}$ hole/atom, (a5) and (b5) $n_{h}=2.5\times 10^{-2}$ hole/atom,  and (a6) and (b6) $n_{h}=3.3\times 10^{-2}$ hole/atom. Band structures (the third column) and corresponding DOS (the fourth column) for bilayer BPN with uniaxial strains and hole doping concentrations: (c1) and (d1) $\varepsilon_{xx}=4.7$ \%, (c2) and (d2) $\varepsilon_{xx}=6.5$ \%, (c3) and (d3) $\varepsilon_{xx}=7.8$ \%, (c4) and (d4) $n_{h}=1.7\times 10^{-2}$ hole/atom, (c5) and (d5) $n_{h}=2.9\times 10^{-2}$ hole/atom, and (c6) and (d6) $n_{h}=3.8\times 10^{-2}$ hole/atom.}
\end{figure*}
Given that the monolayer and bilayer of BPN are metallic, their magnetic behavior can be attributed to the Stoner-type magnetism of itinerant electrons in the metallic system~\cite{Stoner1938, Lidiard_1953, doi:10.1098/rspa.1954.0149, Tredgold_1954, PhysRevB.2.3619, SHIMIZU1984144}. As shown in the band structures, monolayer and bilayer of BPN have distinctive features in the electronic structure: nearly flat bands on $\Gamma X$ and saddle-point vHSs at $\Gamma$ and $Y$. 
These features contribute to an increased DOS that can lead to the Stoner-type magnetism of itinerant electrons. 
To investigate this, we calculate the DOS around $E_{F}$ for monolayer and bilayer of BPN with uniaxial strains and hole doping~\cite{PhysRevB.75.195121}. 

As illustrated in Fig.~\ref{fig:dos}, the DOS exhibits different behaviors when monolayer BPN under uniaxial strains and hole-doped monolayer are compared. 
When uniaxial strains $\varepsilon_{xx}$ are applied, saddle points at $\Gamma$ and $Y$ move close to $E_{F}$, while the nearly flat band shifts away [See Figs.~\ref{fig:dos}(a1)-(a3)]. As a result, the DOS near $E_{F}$ primarily originates from the two saddle points as shown in Figs.~\ref{fig:dos}(b1)-(b3). 
It is known that a saddle point of 2D band structures gives rise to the logarithmically divergent DOS~\cite{PhysRevLett.35.120}. 
As seen in Figs.~\ref{fig:dos}(b1)-(b3), the DOS exhibits two sharp peaks, whose energies correspond to the saddle points at $\Gamma$ and $Y$. 
The peaks have long logarithmic tails.
In contrast, hole-doped monolayer BPN shows enhanced DOS near $E_{F}$ due to the nearly flat band on $\Gamma X$. Since this nearly flat band merges with the saddle point at $\Gamma$, the DOS peak at $E_{F}$ has a logarithmic tail on the lower side of the DOS peak. 
For bilayer BPN, the contributions to the DOS, despite the doubled number of nearly flat bands and saddle points, are similar to monolayer BPN under uniaxial strains and hole doping. 

\begin{figure}[t]
\begin{center}
\includegraphics[width=1.0\columnwidth, clip=true]{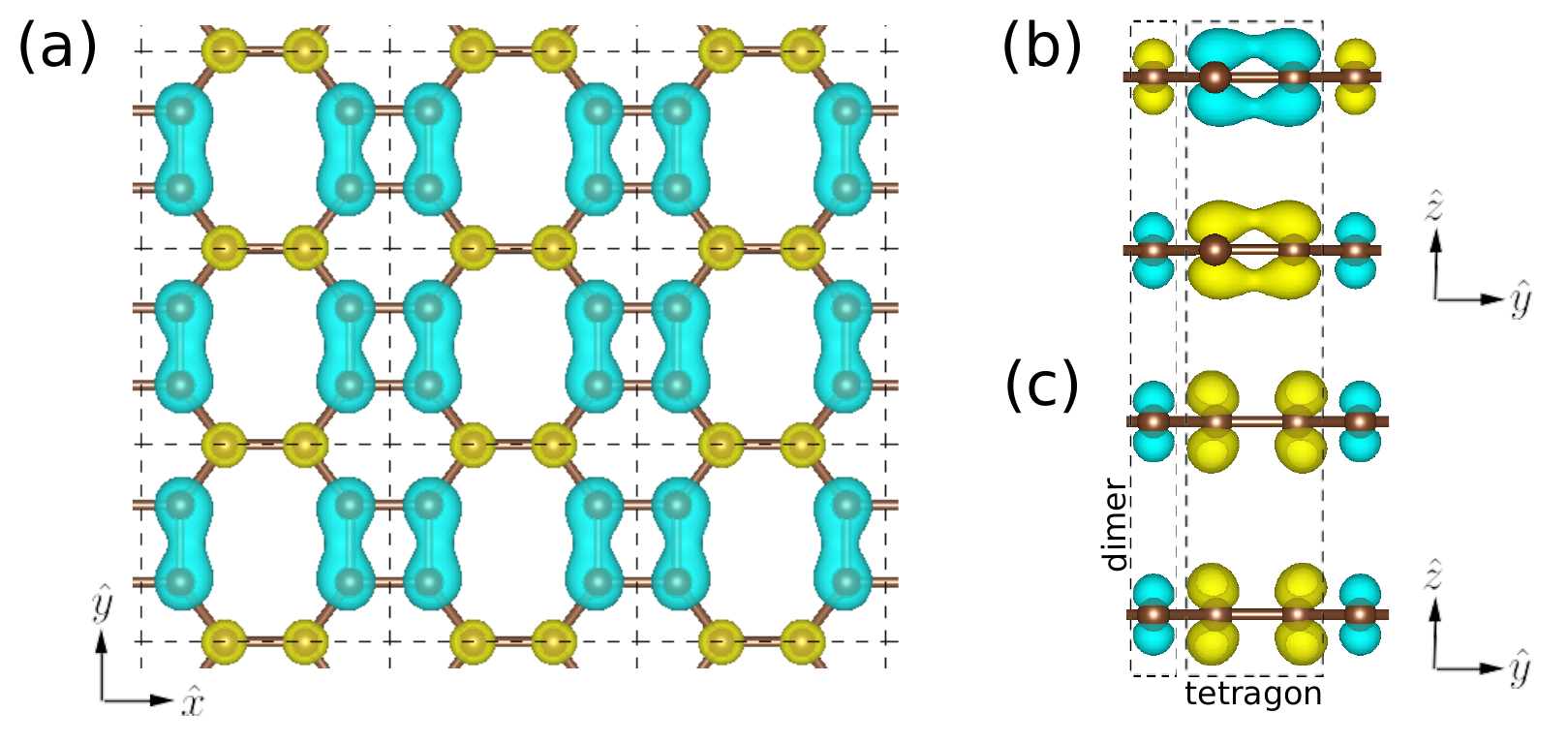} 
\end{center}
\caption{\label{fig:rho} (Color online) Spin polarization density distributions $\Delta \rho = \rho_{\uparrow} - \rho_{\downarrow}$. Cyan and yellow colors represent opposite spins. (a) Top view of spin polarization distribution for single-layer BPN under the $x$ strain of $5.6$ \%. (b) Side view of antiferromagnetic spin polarization distribution when the hole doping is $2.5 \times 10^{-2}$ hole/atom. (c) Side view of ferrimagnetic spin polarization distribution when the hole doping is $4\times 10^{-2}$ hole/atom. Density distributions are drawn by \textsc{VESTA}~\cite{Momma:db5098}}
\end{figure}

\begin{figure}[t]
\begin{center}
\includegraphics[width=1.0\columnwidth, clip=true]{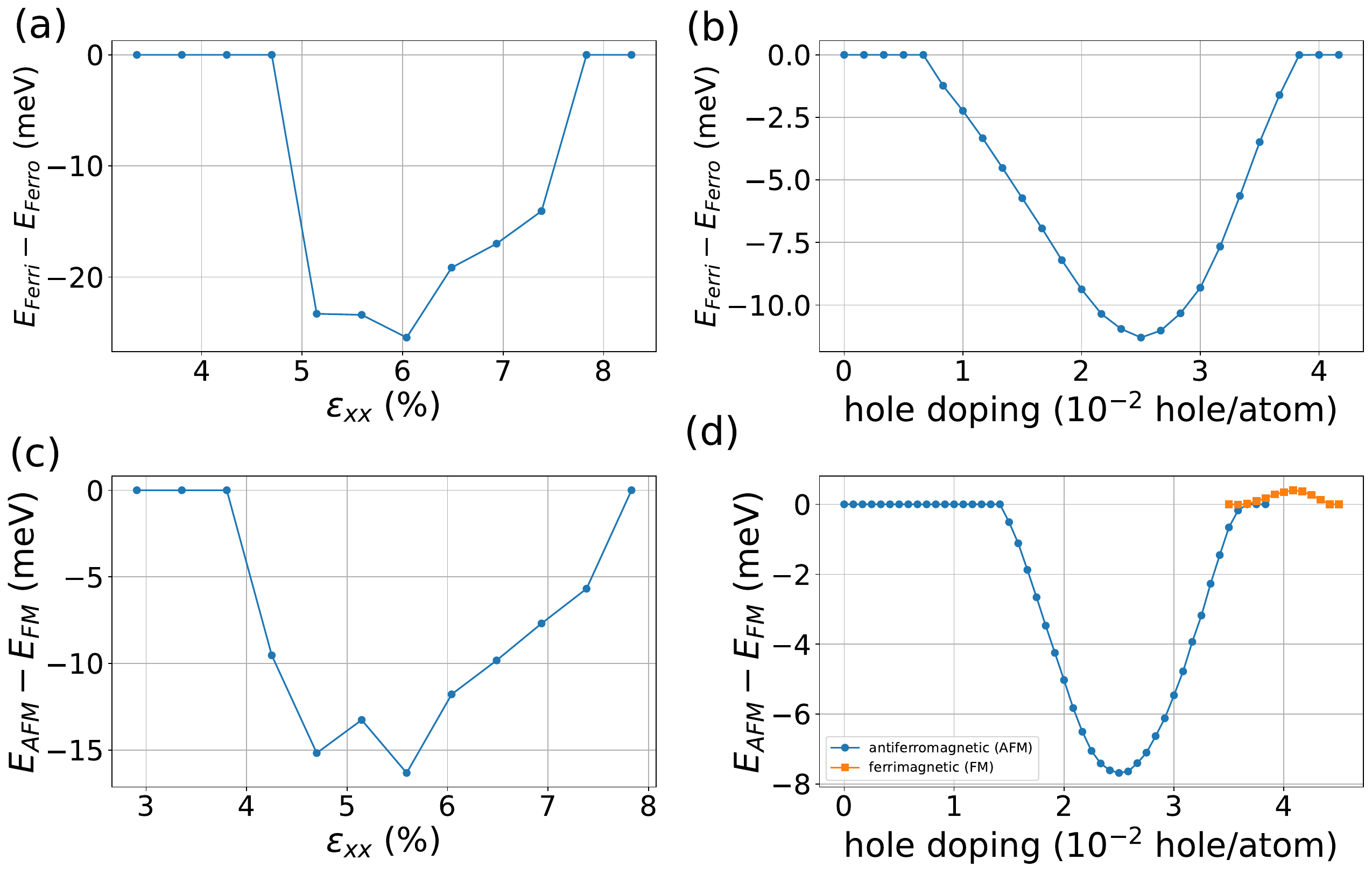} 
\end{center}
\caption{\label{fig:exchange} (Color online) Total energy differences between two magnetic phases of monolayer and bilayer BPN: $E_{\textrm{Ferri}}-E_{\textrm{Ferro}}$ of monolayer BPN as a function of (a) uniaxial strains $\varepsilon_{xx}$ and (b) hole doping concentrations $n_{h}$. $E_{\textrm{AFM}}-E_{\textrm{FM}}$ of bilayer BPN varying (c) $\varepsilon_{xx}$ and (d) $n_{h}$.}
\end{figure}

Lastly, to explore magnetic interactions between magnetic moments, we perform constrained magnetization calculations~\cite{C2DT31662E, PhysRevB.84.224429, PhysRevB.106.195428}.
In this calculation, we maintain the magnitude of magnetic moments, but opposite magnetic moments are flipped so that all magnetic moments are aligned in the same direction. 
As discussed before, monolayer BPN with an appropriate strain $\varepsilon_{xx}$ and hole doping $n_{h}$ displays ferrimagnetic ordering, where majority magnetic moments at the tetragon and minority ones at the dimer unit are aligned in opposite directions. 
For instance, the spin polarization density $\Delta \rho$ at $\varepsilon_{xx}=5.6$ \% are depicted in Fig.~\ref{fig:rho}(a). 
We denote $E_{\textrm{Ferro}}$ as the total energy of the ferromagnetic ordering, where minority magnetic moments at the dimer are flipped in the constrained magnetization calculation. 
Comparing the total energy $E_{\textrm{Ferri}}$ of the ferrimagnetic phase, we calculate the total energy difference $\Delta E = E_{\textrm{Ferri}} - E_{\textrm{Ferro}}$ as a function of $x$-strain $\varepsilon_{xx}$ and hole doping $n_{h}$ in Figs.~\ref{fig:exchange}(a) and (b).
This calculation shows that the transition from the ferrimagnetic phase to the ferromagnetic one requires a positive energy cost, with a maximum of 25 (10) meV, depending on $\varepsilon_{xx}$ ($n_{h}$).    

In the case of bilayer BPN, each layer exhibits ferrimagnetic ordering similar to monolayer BPN. 
In the antiferromagnetic (AFM) phase of bilayer BPN, the lower and upper layers display opposite magnetic moments, as depicted in Fig.~\ref{fig:rho}(b). To be specific, the magnetic moments of the tetragon (dimer) unit in the lower layer have the same magnitude as those in the upper layer, but their directions are opposite. 
We perform constrained magnetization calculations where the magnetic moments of the tetragon (dimer) in the lower and the upper layers are aligned in the same direction and the ferrimagnetic ordering within each layer is fixed, which is referred to as the ferrimagnetic (FM) ordering of bilayer BPN. 
Denoting the resulting total energy as $E_{\textrm{FM}}$, Figs.~\ref{fig:exchange}(c) and (d) illustrate the total energy difference $\Delta E = E_{\textrm{AFM}} - E_{\textrm{FM}}$ for bilayer BPN under uniaxial strains $\varepsilon_{xx}$ and hole doping. 
The flipping of interlayer magnetic moment from the antiferromagnetic phase to the ferrimagnetic one requires an energy cost, suggesting the stability of the antiferromagnetic ordering in bilayer BPN.
In the previous subsection, the highly hole-doped bilayer BPN ($n_{h} > 3.63 \times 10^{-2}$ hole per atom) exhibits ferrimagnetic ordering, as depicted in Fig.~\ref{fig:rho}(c). 
In this case, we compute the total energy of the antiferromagnetic ordering in the constrained magnetization calculations, and compare $E_{\textrm{FM}}$ and $E_\textrm{AFM}$ as shown in Fig.~\ref{fig:exchange}(d). 
For the high hole-doping concentration, the calculation reveals that the ferrimagnetic ordering is energetically more favorable than the antiferromagnetic one, but their total energy difference is less than approximately $0.4$ meV.  

\section{\label{sec:conclusions}Conclusions}
In this study, we examined the magnetic transitions of monolayer and bilayer BPN, under the influence of external perturbations such as uniaxial strains and hole doping.
To account for electron correlations, we employed the DFT+$U$+$V$ method, which incorporates on-site and inter-site interactions as extended Hubbard corrections. 
These extended Hubbard corrections were consistently applied throughout all DFT calculations, encompassing structure relaxations and electronic band structure calculations.

Our findings indicate that both the BPN monolayer and bilayer, in their fully relaxed structures, exhibit non-magnetic behavior. 
However, the application of uniaxial strains along the dimer unit direction induces a ferrimagnetic transition in the monolayer, whereas the bilayer experiences a magnetic transition towards an antiferromagnetic phase.

Furthermore, we explored an alternative approach for inducing magnetic transitions in BPN monolayer and bilayer through hole doping. 
At a moderate level of hole doping, the monolayer assumes a ferrimagnetic phase, while the bilayer adopts an anti-ferrimagnetic state. 
At increased hole doping levels, the bilayer demonstrates the potential for transitioning into a ferrimagnetic phase.
These results offer valuable insights into the intriguing magnetic behavior of BPN-based structures and open up avenues for further investigations in this field.

\section*{Conflicts of interest}
There are no conflicts to declare.

\section*{Acknowledgements}
We thank Jeongwoo Kim, Davide Ceresoli, and Young-Woo Son for fruitful discussions and critical reading. This work was supported by the National Research Foundation of Korea (NRF) grant funded by the Korea government (Grant No.~NRF-2022R1F1A1074670) and by the Open KIAS Center at Korea Institute for Advanced Study.
\bibliography{bibl.bib}
\end{document}